\documentclass[twocolumn,preprintnumbers,amsmath,amssymb]{revtex4}
\usepackage{graphicx}
\usepackage{epsf}

\begin{document}

\title{Density-functional study of the pure and palladium doped small 
       copper and silver clusters}

\author{Hamideh Kahnouji}
\author{Halimeh Najafvandzadeh 
\footnote{H. Kahnouji and H. Najafvandzadeh contributed equally to this work.}}

\author{S. Javad Hashemifar}
\email{hashemifar@cc.iut.ac.ir}

\author{Mojtaba Alaei}
\author{Hadi Akbarzadeh}

\affiliation{Department of Physics, 
             Isfahan University of Technology, 
             84156-83111 Isfahan, Iran}

\begin{abstract}

The size-dependent electronic, structural, magnetic and vibrational   
properties of small pure copper and silver clusters 
and their alloys with one and two palladium atoms are 
studied by using full-potential all-electron density functional computations.
The stable isomers of these clusters are identified and 
their theoretical magic numbers are determined via 
the analysis of the second difference of their minimized energy. 
We discuss that the doped Pd atoms generally prefer to sit in
the high coordination sites of the pure clusters.
It is argued that Pd doping influences the structural properties and the 
two dimensional to three dimensional structural cross over in the small Cu and Ag clusters.
The many body based GW correction is applied for more accurate determination
of the electron affinity and ionization potential of these systems.
Magnetic and vibrational properties of the pure and doped clusters 
are presented and discussed.

\end{abstract}

\maketitle

\section{Introduction}
Noble metal nanoclusters are currently attracting a great deal of attention within both
experimental and theoretical communities, due to their interesting structural
and electronic properties and promising technological applications in nano electronics,
nano optics, biological sensing, catalysis,
and biomedicine \cite{Udayabhaskararao,biological,Burda,Murphy,Xia}.
While transition metal ions may induce health and environmental 
problems \cite{Bioinorgan,Aragay},
noble metal nanoclusters exhibit lower toxicity and hence more biocompatibility.
Noble metal nanoclusters are alloyed with palladium atoms to improve their
catalytic activity \cite{Neerga,WL}.
In recent years, extensive computational studies have been devoted to 
properties of pure noble metal clusters \cite{Berkahem,7,8,9}, 
although less information is available about structural and 
electronic properties of the corresponding bimetallics \cite{10,11,Wissam}.

Among noble metal based bimetallic nanoclusters, copper and silver based nano-alloys
have been studied considerably less than gold based bimetallic nanoclusters.
Wang et al. \cite{Wang} and Romanowski et al.\cite{Roman} used 
density functional (DFT) computations to study AgPd and CuPd dimers 
and their interaction with H$_2$.
Other calculations have been performed on Cu-Pd trimers 
and their interaction with molecular and atomic oxygen \cite{Gobal}.
The stable geometry of small Ag$_n$Pd$_m$ clusters up to five atoms 
were found to transform from two-dimensional 
to three-dimensional as the Pd content increases \cite{Papageo}.
Zhao et al. performed first-principles calculations to study 
the effects of Pd doping on structural and electronic properties 
of Ag$_n$Pd ($n\leq5$) clusters and their mono hydrides \cite{Zhao}.
Efremenko et al.\cite{32} have studied the geometric structures and electronic 
properties of the Pd$_n$Cu$_m$ ($n+m\leq6$) clusters using DFT. 
The obtained results show that stability of nanoclusters of the same shape 
and composition increases linearly with increasing number of Pd–Cu bonds.
Up to our knowledge, there is no report on magic number, vibrational spectra,
and many body corrected electronic structure of small Pd doped Ag and Cu clusters.

In this paper, quantum mechanical calculations are employed to study 
pure M$_n$ (M = Ag, Cu) and doped M$_{n-m}$Pd$_m$ ($n\leq9, m=1,2$) clusters 
and investigate the behavior of their structural, electronic, and vibrational
properties as a function of size.
The rest of the paper is organized as follows. 
In section II, we give a brief introduction of the computational method 
used in this work.
Then in section III, atomic dimers are investigated with several  
exchange-correlation functional to select the proper one for our calculations.
Next section involves the results of our search for lower energy structures
of the selected nano-clusters.
In section IV, the structural, electronic, magnetic and vibrational properties 
of the most stable clusters are presented. 
Finally, we  will summarized our findings in the last section.

\section{Method}

All presented calculations are performed in the framework of 
the spin-polarized Kohn-Sham density functional theory by using the all electron 
full-potential code {\sc FHI-aims} \cite{FHI-aims}. 
This package employs basis sets consisting of atom-centered 
numerical orbitals of the form:
\begin{equation}
 R(r)=\dfrac{u_{i}(r)}{r}
\end{equation}
As the name implies, the radial free atom like orbitals $u_i(r)$
are numerically tabulated and therefore exhibit very high flexibility. 
The strict localized nature of these basis functions lead to much slower
scaling of the computational time versus the system size.
The calculations reported here are done with the "tier2+spd", "tier1+spd",
and "tier2" basis sets which contain 116, 152, and 45 basis functions 
for Cu, Ag, and Pd atoms, respectively.
All the calculations were carried out in the scalar relativistic limit,
while the spin-orbit correction was neglected.
A recent computational study on the atomic structure of transition metal clusters
concluded that this relativistic term has negligible effect on the relative energy
differences of the 4d systems \cite{17a}.

Geometry optimization is performed by the standard Broyden-Fletcher-Goldfarb-Shanno
(BFGS) algorithm \cite{19} with a force accuracy of about $10^{-2} eV/\AA$. 
Harmonic frequencies are calculated using
the finite displacement of all atomic positions by $10^{-3} \AA$. 
The lowest energy structures are confirmed to be the true minima by calculating 
their vibrational frequencies.
The FHI-aims package enables us to describe electronic single-quasiparticle 
excitations in molecules by using many-body correction GW self-energy \cite{gw}.

\section{Dimers}

First we focus on the atomic dimers to select a reliable exchange-correlation 
(XC) functional for our computations and moreover to gain some insights
about interatomic interaction in our systems. 
The equilibrium bond length, binding energy, and 
vibrational frequency of the Cu$_2$ and Ag$_2$ dimers were calculated 
by using different XC approximations, including PBE \cite{20}, 
revPBE \cite{21} and BLYP \cite{blyp} generalized gradient (GGA) functionals 
and PW \cite{pw} local density functional.  
The obtained results are compared with available experimental data 
in table \ref{dimer}. 
It is seen that PBE, revPBE, and BLYP give significantly more accurate
binding energy and vibrational frequency while accuracy of
equilibrium bond length is qualitatively the same within LDA and
GGA based functionals.
In the case of Cu$_2$, the overall accuracy of revPBE seems to be slightly
better than PBE, while the parameters of Ag$_2$ within PBE is overall
closer to experiment. 
Although, BLYP gives better binding energy for Ag dimer, for larger silver clusters, 
other theoretical calculations \cite{27} argue that the BLYP functional 
fails to predict the correct stable structure of the system.
Therefore, we adapt the PBE and revPBE functionals for 
calculation of Ag and Cu based clusters, respectively.

\begin{table}
\newcommand{\ra}{\cite{22}}
\newcommand{\rb}{\cite{23}}
\newcommand{\rc}{\cite{24}}
\newcommand{\rd}{\cite{25}}
\newcommand{\re}{\cite{28}}
\newcommand{\rf}{\cite{29}}
\newcommand{\rh}{\cite{30}}
\caption{\label{dimer}
 Calculated equilibrium bond length $r_e$ (\AA), 
 binding energy $E_b$ (eV/atom) and harmonic vibrational frequency 
 $\omega$ (cm$^{-1}$) of the Cu$_2$, Ag$_2$, Pd$_2$, CuPd, and AgPd dimers.
 The Cu$_2$ and Ag$_2$ dimers are calculated in four different 
 exchange-correlation (XC) functionals.
 The corresponding experimental values (Expt.) are 
 also given for comparison.
}
\begin{ruledtabular}
\begin{tabular}{lllll}
       &  XC    & $r_e$    &  $E_b$    & $\omega$ \\
\hline
Cu$_2$ & PBE    & 2.21     & -1.13     & 269     \\
       & BLYP   & 2.23     & -1.11     & 261     \\
       & revPBE & 2.24     & -1.02     & 259     \\
       & PW     & 2.15     & -1.39     & 299     \\
       & Expt.  & 2.21~\ra & -1.04~\rb & 265~\rb \\
Ag$_2$ & PBE    & 2.57     & -0.91     & 180     \\
       & BLYP   & 2.60     & -0.83     & 172     \\
       & revPBE & 2.60     & -0.79     & 171     \\
       & PW     & 2.48     & -1.45     & 270     \\
       & Expt.  & 2.53~\rc & -0.83~\rd & 192~\rc \\
CuPd   & revPBE & 2.32     & -0.79     & 232     \\    
AgPd   & PBE    & 2.56     & -0.70     & 179     \\    
Pd$_2$ & PBE    & 2.49     & -0.49     & 195     \\  
       & Expt.  & 2.48~\re & -0.51~\rf & 210~\rh \\  
\end{tabular}
\end{ruledtabular}
\end{table}

For better understanding of the interatomic bonds in our target systems,
we calculated the Pd$_2$, CuPd, and AgPd dimers and 
presented their results in table \ref{dimer}.
It is observed that the bond length is increasing in 
the Cu$_2$ - CuPd - Pd$_2$ series while the absolute binding energy and 
vibrational frequency is decreasing.
Hence, Cu-Cu bond has a larger strength and stiffness compared
with the Cu-Pd and Pd-Pd bonds.
Bond length and normal mode frequencies of AgPd and Ag$_2$ dimers 
are almost equal while these are larger than those of Pd$_2$ dimer.
The larger bond length of Ag$_2$ may be due to the 
large atomic radius of Ag relative to Pd and Cu atoms.
Binding energy of Pd$_2$ is smaller than that of AgPd 
which in turn is smaller than Ag$_2$ dimer.

The Mulliken population analysis of valence orbitals shows that 
Cu and Ag have a similar electronic configuration of $s^{0.94}p^{0.05}d^{10.01}$
in the Cu$_2$ and Ag$_2$ dimers.
Comparing this configuration with the free Cu and Ag atoms
electronic configuration ($s^{1}p^{0}d^{10}$) evidences 
a small $spd$ hybridization in these dimers.
On the other hand, comparing the electronic configuration of Pd in 
the Pd$_2$ dimer ($s^{0.58}p^{0.05}d^{9.37}$) with 
the free Pd atom ($s^{0}p^{0}d^{10}$) indicates
a significant $spd$ hybridization in the palladium dimer.
The reason is that the valence shell of the free Pd atom is composed
of fully occupied and fully unoccupied orbitals and hence interatomic
bonding between palladium atoms requires significant promotion
of electrons from the occupied $d$ states to the unoccupied $s$ shell.
While, the valence shell of the free Cu and Ag atoms has a half filled 
$s$ orbital which is used for interatomic bonding.
It is also found that the charge transfer in the Pd-Cu and Pd-Ag bonds 
happens from Cu and Ag to the more electronegative Pd atom.

\section{Stable isomers}

\begin{figure}
\includegraphics*[scale=0.5]{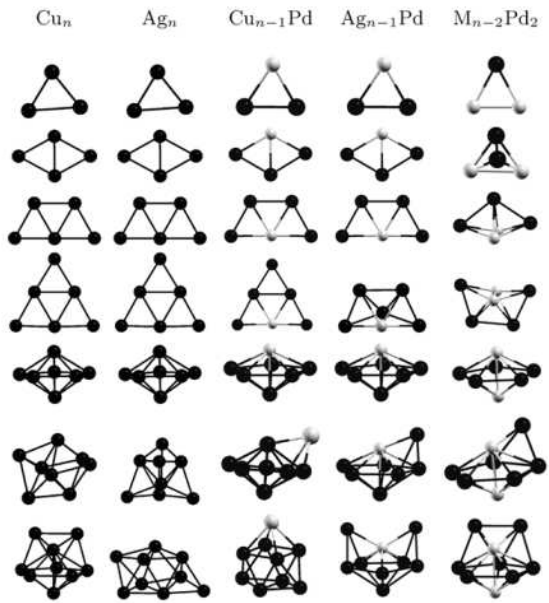}
\caption{\label{isomer}
 Obtained lowest energy isomers of the pure Ag$_n$ and Cu$_n$
 and doped Cu$_{n-1}$Pd, Ag$_{n-1}$Pd, Cu$_{n-1}$Pd$_2$, 
 and Ag$_{n-1}$Pd$_2$ clusters for $3\leq n\leq9$. 
 The Pd atoms are shown by light balls while the Cu/Ag atoms 
 are indicated by black balls.
}
\end{figure}

In order to investigate properties of atomic clusters, the first essential step
is identification of the lowest energy structure of the clusters.
Therefore, we performed a careful search for the stable structure of the pure 
and doped clusters. 
In the case of the pure clusters, all probable atomic configurations of 
the clusters were included in our search. 
After accurate atomic relaxation of the relevant configurations and comparing
their minimized total energies, the most stable isomers of the pure 
Cu$_n$ and Ag$_n$ clusters ($n\le9$) were identified and sketched in Fig.~\ref{isomer}.
These findings are in agreement with a recent study on silver clusters \cite{31}.
Because of the presence of two different atoms, the process of 
finding the global minimum energy structure of the doped clusters 
is more complicated than the pure clusters.
Therefore, for finding the most stable structures of the palladium doped clusters, 
we tried to limit our search to the more stable configurations
proposed by previous studies on the Cu$_n$Pd$_m$ ($n+m\leq6$)\cite{32}
and Ag$_n$Pd$_m$ ($n+m\leq5$)\cite{33,Papageo} clusters.
The resulting most stable isomer of the doped clusters are 
presented in Fig.~\ref{isomer}.
It is seen that the most stable M$_{n-2}$Pd$_2$ clusters are 
made of the maximum number of pyramids and moreover
the two Pd atoms in these systems tend to bond together.
These observations were used for obtaining the most stable 
structure of the M$_6$Pd$_2$ and M$_7$Pd$_2$ clusters.

\begin{figure}
\includegraphics*[scale=0.95]{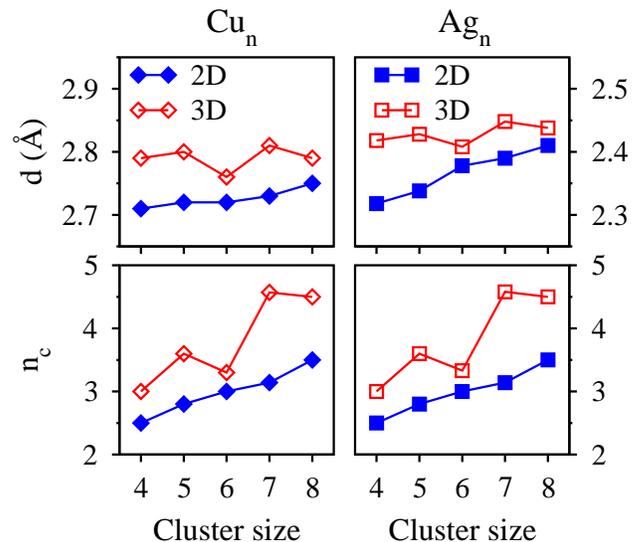}
\caption{\label{cross}
 Average bond length $d$ and coordination number of atoms $n_c$ 
 in the most stable 2D and 3D structures of Cu$_n$ (left) and 
 Ag$_n$ (right) clusters.
}
\end{figure}

The results show that up to the size of 6 (heptamer), pure clusters prefer 
2 dimensional (2D) planar structures while larger clusters 
stabilize in 3D geometries. 
In order to understand the origin of this 2D-3D cross over, 
we analyzed the average bond length ($d$) and the average 
coordination number of atoms ($n_c$) in the most stable 2D and 3D
isomers of the Cu$_n$ and Ag$_n$ clusters (Fig.~\ref{cross}).
It is seen that the 2D isomers have lower average coordination and
shorter average bond length, because the valence charge density of
these systems is distributed among lower number of bonds,
compared with the 3D isomers.
As a result of that, individual bonds in the planner isomers are stronger and shorter,
while 3D isomers have more number of bonds.
The 2D-3D structural transformation reflects the competition between
the individual bond strength and the total number of bonds.
It is observed that (Fig.~\ref{cross}) in the smaller sizes ($n\le6$),
the more pronounced difference in the individual bond strength leads to 
the stability of the 2D isomers, while for the larger clusters 
the difference in the average coordination wins the competition and 
stabilizes the 3D isomers.

The results (Fig.~\ref{isomer}) indicate that the most stable 2D isomers of 
Cu$_n$ and Ag$_n$ clusters are made of equilateral triangles. 
It is also seen that the 2D-3D cross over in the Ag$_{n-1}$Pd    
clusters occurs at $n=6$ while the Cu$_{n-1}$Pd clusters,
follow the same structural transformation as the pure Cu cluster.
As a general trend, we observe that the Pd atom prefers the high coordinated
sites of the Cu$_{n-1}$Pd and Ag$_{n-1}$Pd clusters.
The obtained results indicate high tendency of the M$_{n-2}$Pd$_2$ 
clusters toward 3D configurations. 
Considering various 2D and 3D structures for the M$_2$Pd$_2$, it was found that
the 3D structures have lower energies, in agreement with the previous works \cite{32,33}.
Based on these results, the theoretical search for the stable isomers of 
the M$_{n-2}$Pd$_2$ clusters were limited to all 3D isomers of the pure Ag and Cu clusters. 
It is worth while to mention that other authors have not investigated
Cu$_{n-2}$Pd$_2$ and Ag$_{n-2}$Pd$_2$ clusters larger 
than five and six atoms, respectively.

\section{Structural properties}

\begin{figure}
\includegraphics*[scale=0.95]{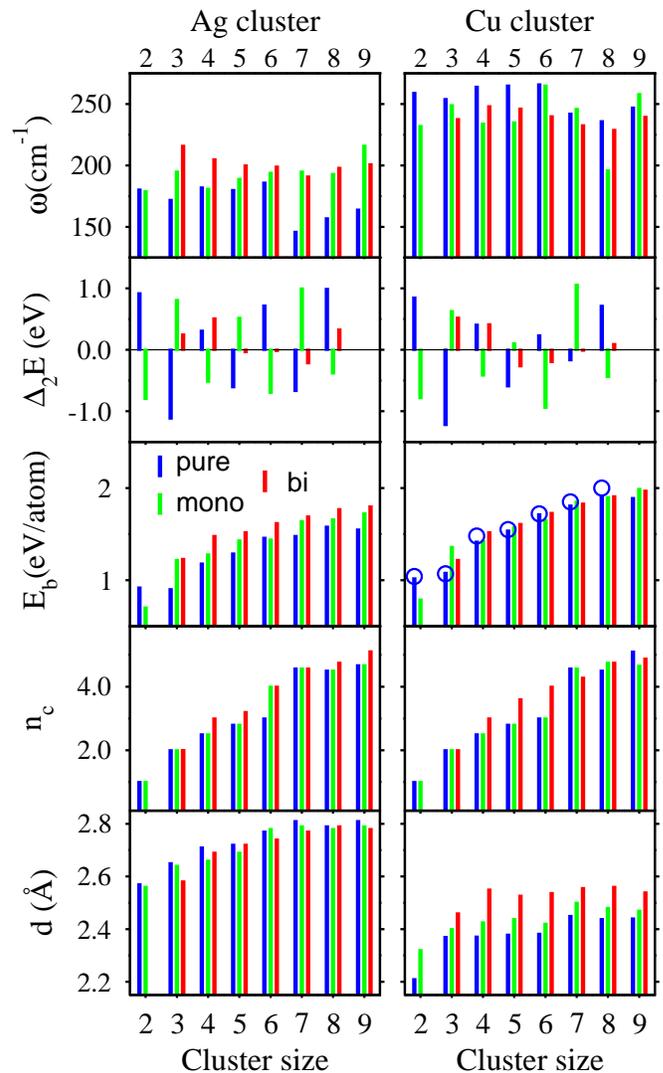}
\caption{\label{bond}
 Calculated maximum vibrational frequency ($\omega$),
 Second difference of the energy $\Delta_{2}E(n)$,
 absolute binding energy per atom $E_b$, 
 average coordination number of atoms $n_c$, 
 and the average bond length $d$ of the pure,
 mono-doped, bi-doped Ag and Cu clusters
 as a function of size.
 The empty circles show the measured binding energy 
 of the pure Cu clusters \cite{Eb-Cu}.
 }
\end{figure}

In order to address stability and structural behaviour of the clusters,
the binding energy per atom, average coordination number of atoms,
and the average bond length of the most stable structures of all clusters 
under study as a function of the cluster size are presented in Fig. \ref{bond}.
The harmonic vibrational frequencies of the clusters were also calculated 
to address dynamical stability and bond stiffness of the systems.
The maximum vibrational frequency of the clusters are shown in Fig. \ref{bond}.
The calculated binding energy of the copper clusters are in well agreement
with the measured data.
It can be seen that by increasing the cluster size, 
the binding energy of the pure clusters increases gradually toward 
the binding energy of bulk copper (3.50 eV)\cite{35} and silver (2.95 eV)\cite{35},
although these clusters are clearly far from the bulk limit. 
Notably, the eight atom pure clusters occurs in a local maximum 
of the absolute binding energy. 
It is clearly seen that the 2D to 3D structural transformation in 
the pure and mono-doped clusters is reasonably accompanied by a sharp 
increases in the average coordination number of atoms.

Observation of no imaginary frequency indicates the dynamical stability of
the lowest energy isomers shown in Fig. \ref{isomer}.
It is generally seen that the Ag based clusters have lower 
vibrational frequency than Cu based clusters, which is clearly
related to the higher atomic mass of silver.
Considering the trend of maximum frequencies along with the geometry of the most stable 
isomers of the clusters (Fig. \ref{isomer}) clarifies that the structural cross over
from planner to 3D geometries in the pure Ag and Cu clusters and doped Cu$_{n-1}$Pd cluster
is accompanied with a significant mode softening in the vibrational spectra.

Comparing the properties of the pure and doped clusters show that 
Pd doping in the copper clusters increases the average 
bond length of the system, because of the larger atomic radius 
of Pd compared with Cu.
This increase is expected to be accompanied by a bond softening effect,
which is generally visible in the calculated harmonic vibrational 
frequency of the doped Cu clusters.
In contrast to this bond elongating and softening effects,
we observe that Pd doping has a week effect on the binding energy of Cu clusters.
It seems that the doped Pd atom in the Cu cluster, inserts effective
amount of electrons in the neighboring bonds and hence compensate
the observed bond softening effects.
It is reasonable, because Pd has much more electrons than Cu.
On the other hand, the average bond length of the doped Ag clusters 
is slightly smaller than those of the pure clusters. 
The reason is that atomic size of silver is larger than that of palladium 
and hence Ag-Pd and Pd-Pd bonds are shorter than Ag-Ag one (table~\ref{dimer}).
As a result of that, a bond hardening effect is visible in the calculated
vibrational spectra of the doped Ag clusters, compared with the pure ones.
As a result of these bond shortening and hardening effects,
Pd doping enhances the absolute binding energy of Ag clusters (Fig.~\ref{bond}).

In cluster science, the second-order energy difference ($\Delta_{2}E$) 
is a conventional quantity to address the relative stability of clusters
as a function of size and reproduce the experimental magic numbers
observed in mass spectra measurements. 
Large values of $\Delta_{2}E(n)$ indicate that the $n$-atoms cluster 
is more stable than $n-1$ and $n+1$ atoms clusters. 
This parameter is defined as follows:
\begin{equation}
\Delta_{2}E(n) = E_{tot}(n+1) + E_{tot}(n-1) - 2 E_{tot}(n)
\end{equation}
$E_{tot}(n)$ is the minimized energy of the M$_{n-m}$Pd$_m$ cluster
(M = Ag, Cu) and $E_{tot}(n+1)$ ($E_{tot}(n-1)$) is the minimized 
energy of the cluster with one more (less) M atom. 
The calculated second-order energy difference of the pure and 
doped Cu and Ag clusters is plotted in Fig. \ref{bond}.
It is seen that pure clusters show strong peaks at $n=2,8$ which 
indicate the highest relative stability of these systems and
propose $n=2,8$ as the first magic numbers of small pure Ag and Cu clusters,
in agreement with the experimental observations
\cite{exper1,exper2,exper3,exper4} 
and electronic shell jellium model \cite{jellium1,jellium2}. 
The predicted magic numbers for mono-doped clusters are 3 and 7,
while in the case of bi-doped clusters, the main peaks are observed at $n=3,4,8$.

Furthermore, a clear odd-even oscillation is seen in the pure
and mono-doped clusters, although bi-doped clusters
do not display such clear oscillations. 
It is worthwhile to mention that electron pairing in HOMO can explain this
oscillations and the more stability of the even pure and odd mono-doped clusters.
The valence electrons of the pure M$_n$ and mono-doped M$_{n-1}$Pd clusters
are mainly composed of the M $s$ electrons, 
hence clusters with odd number of M atoms have one unpaired electron 
and consequently exhibit lower stability.
The pure and mono-doped clusters with even number of M atoms have 
a close electronic shell and hence are more stable. 
On the other hand, the behaviour of the $\Delta_{2}E$ curve of the bi-doped clusters
is more complicated and needs to further arguments.
We attribute this more complicated behaviour to the effective contribution
of $d$ electrons to the valence shell of the bi-doped clusters.
As it was argued in section III,
the direct bonding between Pd atoms in the bi-doped clusters
give rises to effective promotion of the Pd d electrons to 
the valence shell of the system.

\begin{figure}
\includegraphics*[scale=0.93]{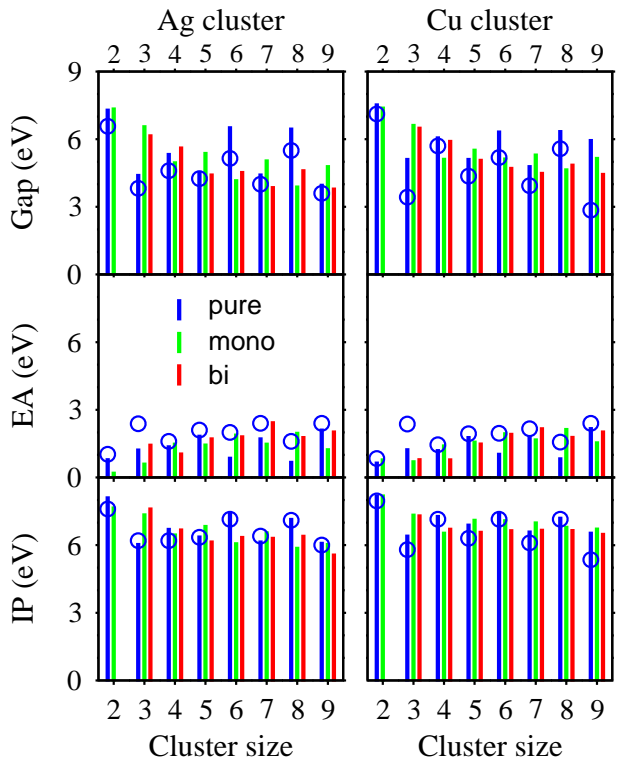}
\caption{\label{band}
 Calculated HOMO-LUMO gap, Ionization potential (IP), 
 and electron affinity (EA)
 of the pure, mono-doped, and bi-doped
 Ag (left column) and Cu clusters (right column) after the GW correction.
 The empty circles show the available measured data
 for pure clusters, collected from references 
 \cite{Ip-Ag, EA-Ag, EA-Cu1, EA-Cu2, IP-Cu1,IP-Cu2}.
}
\end{figure}

\section{Electronic properties}

\begin{figure}
\includegraphics*[scale=0.93]{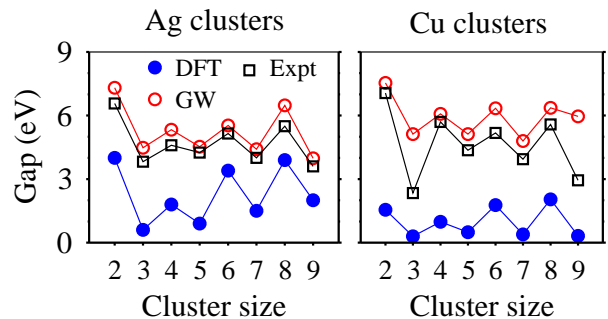}
\caption{\label{gap}
 Calculated HOMO-LUMO gap of the pure Ag and Cu clusters
 before and after the GW correction
 along with the available experimental data,
 collected from references
 \cite{Ip-Ag, EA-Ag, EA-Cu1, EA-Cu2, IP-Cu1,IP-Cu2}.
}
\end{figure}

Since the single particle Kohn-Sham eigenvalues have no clear physical meaning, 
the reliable energy level of the highest occupied (HOMO) and the lowest 
unoccupied molecular orbital (LUMO) of the clusters are determined 
after applying the many-body GW perturbation theory \cite{gw}.
The GW corrected HOMO and LUMO energies are expected to be 
comparable to the experimental ionization potential (IP) 
and electron affinity (EA) of the clusters.
These values along with the computed HOMO-LUMO gaps are presented in Fig.~\ref{band}.
The available measured data for pure clusters are also presented in this figure.
The calculated values of IP and Gap exhibit good agreement with 
the experimental data which is mainly due to the GW correction,
while the calculated EA values in some cases are considerably overestimated.
In order to show the effect of this many body correction,
we have compared in Fig. \ref{gap} the experimental values of the HOMO-LUMO gap 
of the pure clusters with the calculated ones before and after 
application of the GW correction.
It is seen that the Kohn-Sham gaps are significantly underestimated
while the GW corrected gaps are generally much closer to experiment.
The observed odd-even oscillations in the IP and EA of the pure clusters are 
attributed to the electron pairing in these systems, discussed in the previous section.
The lower value of EA for even clusters indicates less tendency of these closed shell systems 
to receive an extra electron, compared with the open shell clusters.
A consistent reversed trend is visible in the obtained ionization potentials.

In the same way, the ionization potential and electron affinity of the most 
stable structure of the doped clusters were calculated and displayed in Fig.~\ref{band}. 
The higher value of IP for the odd M$_{n-1}$Pd clusters indicates their more 
chemical stability compared to the neighboring even clusters.
In the bi-doped clusters, the IP and EA curves do not show clear oscillations 
with respect to cluster sizes, which is likely due to the effective
contribution of $d$ electrons to the Pd-Pd bonding in these clusters.  
The trend of the calculated HOMO-LUMO gap is generally consistent with 
the $\Delta_{2}E$, indicating that the systems with higher gap are
generally more stable.
It is consistent with the chemical intuition that a higher value of 
the HOMO-LUMO gap indicates less chemical activity and hence more stability.
Considering the spin polarized electronic structure of the most stable isomers,
we found out that consistent with the simple spin pairing rule,
all studied clusters prefer the lowest possible spin multiplicity.
Hence, the pure and doped clusters with even number of Ag and Cu atoms
are nonmagnetic while others exhibit a total spin moment of 1 $\mu_B$.

\section{Conclusions}

In this paper, spin-polarized all-electron calculations were performed 
to investigate structural and electronic properties of the pure
and Pd doped Cu$_n$ and Ag$_n$ clusters ($n\le9$). 
It was argued that a 2D to 3D structural cross over occurs in the size
of seven of the pure clusters. The magic numbers of the pure clusters 
were found to be 2 and 8.
The obtained results indicate that the doped Pd atoms
generally prefer the high coordinated sites of the pure clusters.
We found that the magic numbers of the mono-doped clusters are 
3 and 7 while bi-doped clusters exhibit magic numbers 3, 4, and 8. 
It was argued that doping with one Pd atom slightly reduces the onset
of the 2D to 3D structural crossover of the pure Ag clusters,
while doping with two Pd atoms strongly enhances stability
of the 3D isomers and completely destroys this cross over.
Moreover, it was observed that the bond stiffness of the Cu clusters 
are decreased after doping with Pd atoms while their 
average bond strength have lower sensitivity to doping.
On the other hand, Pd doping increases the bond stiffness 
and bond strength of the Ag clusters.
It was discussed that electron pairing in the HOMO,
induces a significant odd-even fluctuation in the ionization potential,
electron affinity, and HOMO-LUMO gap of the pure Ag and Cu clusters,
while doping with Pd atoms considerably weakens this fluctuation.

\section*{Acknowledgments}

This work was jointly supported by the Vice Chancellor
for Research Affairs of Isfahan University of Technology (IUT),
Centre of Excellence for Applied Nanotechnology,
and ICTP Affiliated Centre at IUT.

\bibliography{ag-cu}

\end{document}